\documentclass[runningheads]{llncs}

\usepackage{fancyhdr}

\fancypagestyle{specialfooter}{%
  \fancyhf{}
  
  \fancyfoot[C]{\vspace{1cm} \hspace*{-.2\textwidth}\parbox{1.4\textwidth}{\small This is the author's version of the work. It is posted here for your personal use. The definitive version is published in: \\ \emph{Proceedings of the 45th European Conference on Information Retrieval} (ECIR '23), April 2--6, 2023, Dublin, Ireland}}
}

\usepackage{url}
\usepackage{tabularx}
\usepackage{multirow}
\usepackage{color}
\usepackage{booktabs}
\usepackage[square,comma,numbers,sort&compress,sectionbib]{natbib}
\usepackage{enumerate}
\usepackage{graphicx}
\usepackage{subfig}
\usepackage{diagbox}

\makeatletter
\newcommand{\speciallabel}[2]{
  \edef\@currentlabel{#1}\label{#2}%
}
\def\@xfootnote[#1]{%
  \protected@xdef\@thefnmark{#1}%
  \@footnotemark\@footnotetext}
\makeatother

\begin{document}
\mainmatter

\title{From Baseline to Top Performer: \\A Reproducibility Study of Approaches at the TREC 2021 Conversational Assistance Track}

\titlerunning{From Baseline to Top Performer}  

\author{Weronika Lajewska \and \mbox{Krisztian Balog}}
\institute{University of Stavanger, Stavanger, Norway,\\
\email{\{weronika.lajewska, krisztian.balog\}@uis.no}}

\authorrunning{Lajewska and Balog} 
\tocauthor{Weronika Lajewska, Krisztian Balog}

\maketitle

\thispagestyle{specialfooter}

\begin{abstract}
This paper reports on an effort of reproducing the organizers' baseline as well as the top performing participant submission at the 2021 edition of the TREC Conversational Assistance track. TREC systems are commonly regarded as reference points for effectiveness comparison. Yet, the papers accompanying them have less strict requirements than peer-reviewed publications, which can make reproducibility challenging. Our results indicate that key practical information is indeed missing. While the results can be reproduced within a 19\% relative margin with respect to the main evaluation measure, the relative difference between the baseline and the top performing approach shrinks from the reported 18\% to 5\%. Additionally, we report on a new set of experiments aimed at understanding the impact of various pipeline components. We show that end-to-end system performance can indeed benefit from advanced retrieval techniques in either stage of a two-stage retrieval pipeline.  We also measure the impact of the dataset used for fine-tuning the query rewriter and find that employing different query rewriting methods in different stages of the retrieval pipeline might be beneficial. Moreover, these results are shown to generalize across the 2020 and 2021 editions of the track. We conclude our study with a list of lessons learned and practical suggestions.

\keywords{Conversational search \and Query rewriting \and Dense retrieval \and Passage re-ranking \and TREC CAsT \and Reproducibility}
\end{abstract}

\section{Introduction}
The last few years have seen an acceleration of research on multi-turn, natural language, and long-term user modeling capabilities of search systems with an attempt to make them more conversational~\citep{Zamani:2022:arXiv}. 
The Conversational Assistance Track (CAsT) at the Text Retrieval Conference (TREC)~\citep{Dalton:2019:TREC,Dalton:2020:TREC,Dalton:2021:TREC} has been a key enabler of progress in this area, by providing a reusable test collection for conversational search. 
The task at TREC CAsT is to identify relevant content from a collection of passages, ``for conversational queries that evolve through a trajectory of a discussion on a topic''~\citep{Dalton:2021:TREC}.
Over the years, query rewriting, passage retrieval, and passage reranking have emerged as the main components, which are combined in a pipeline architecture.  Clearly, the ranking components can directly benefit from advances in dense/hybrid passage retrieval~\citep{Luan:2021:MIT}, and are indeed critical to overall system performance.  However, what makes the task interesting from a conversational perspective, and different from passage retrieval, is the problem of query rewriting~\citep{Kumar:2020:EMNLP,Lin:2021:TOIS,Vakulenko:2021:ECIR,Vakulenko:2021:WSDM,Yu:2020:SIGIR,Voskarides:2020:SIGIR,Mele:2020:SIGIR}.

It has been shown that the best performing systems at TREC form a very competitive reference point for effectiveness comparison~\citep{Armstrong:2009:CIKM}.  
This means that, even if one's ultimate research interest lies in query rewriting, demonstrating strong absolute performance for conversational search requires a high degree of effectiveness from all system components.
Our main objective in this paper is to reproduce (1) the best performing baseline method provided by the track organizers~\citep{Dalton:2021:TREC} and (2) the top performing (documented) system~\citep{Yan:2021:TREC} from the latest (2021) edition of TREC CAsT. 
These two approaches are seen as representatives of a strong baseline and the state of the art, respectively. 
It is worth noting that the system description papers accompanying TREC submissions are not peer-reviewed and there is no explicit or implicit reproducibility requirement.  This can make reproducibility particularly challenging and a study such as this particularly insightful.

Both selected systems follow a two-stage \emph{retrieve-then-rerank} pipeline architecture with queries rewritten based on conversational context.
Specifically, the baseline system~\citep{Dalton:2021:TREC} uses a T5-based query rewriting model fine-tuned on CANARD~\citep{Elgohary:2020:EMNLP-IJCNLP}, first-pass retrieval based on BM25, and a pointwise T5 re-ranker.
The top participating system~\citep{Yan:2021:TREC} uses a different dataset for fine-tuning the query rewriting model (QReCC~\citep{Anantha:2021:NAACL}) and employs more advanced ranking components: a combination of sparse-dense retrieval with pseudo relevance feedback for first-pass retrieval (ANCE/BM25/PRF), and pointwise/pairwise (mono/duoT5) re-ranking.
We find that these complex multi-step architectures are challenging to reproduce due to the numerous components involved.  
Neither the baseline nor the top participating system can be fully reproduced due to key information missing about model choices, parameters, and various input preparation and collection preprocessing steps.
With a best-effort attempt, the results we obtain are within 12\% and 21\% relative margins for the baseline and top performing systems, respectively, with regards to all evaluation measures reported in the track overview paper~\citep{Dalton:2021:TREC}.
However, the relative differences between the two shrink from 18\% in NDCG@3 and 37\% in recall, according to the track overview, to 5\% and 7\%, respectively, according to our reproduced systems.

Since the two selected systems follow the same basic two-stage retrieval pipeline, we perform additional experiments in order to better understand how each pipeline component contributes to overall effectiveness. To shed light on the generalizability of findings, we report results on both the 2020 and 2021 editions of TREC CAsT.  
Since the query rewriter influences the effectiveness of both first-pass retrieval and re-ranking, we also perform experiments using a different retrieval pipeline, which can utilize different query rewriting methods for the two ranking stages. 
We find that final performance is indeed influenced by the position of the query rewriting component in the retrieval pipeline: T5 fine-tuned on CANARD gives better results than fine-tuning on QReCC in terms of first-pass retrieval (higher recall), whereas the best overall results (NDCG@3) are achieved by the system using QReCC for first-pass retrieval and CANARD for re-ranking.
This suggests that employing different query rewriting methods for the different stages might be beneficial.

In summary, the main contributions of this paper are twofold.
First, we attempt to reproduce two approaches from the latest edition of TREC CAsT, the organizers' baseline and the top performing submission, and report results and lessons learned.
Second, we present additional experiments on two-stage retrieval pipelines and query rewriting models to provide insights into the potential contributions of various components.
All resources developed within this study (i.e., source code, runfiles, evaluation results) are made publicly available.\footnote{\url{https://github.com/iai-group/ecir2023-reproducibility}}

\section{Related Work}
\label{sec:related-work}

We briefly introduce the TREC Conversational Assistance Track, discuss query rewriting approaches, and review ranking architectures used at TREC CAsT.

\subsection{TREC Conversational Assistance Track}

The Conversational Assistance Track at TREC has started in 2019 with the aim to facilitate research on conversational information seeking, by creating a large-scale reusable test collection~\citep{Dalton:2019:TREC}. 
The task is to identify relevant passages (in 2019 and 2020) or documents (in 2021) from a collection comprising MS MARCO~\citep{Campos:2016:arXiv},
Wikipedia~\citep{Petroni:2020:arXiv}, 
TREC CAR~\citep{Dietz:2018:TREC} 
and the Washington Post v4.\footnote{\url{https://trec.nist.gov/data/wapost/}}

In TREC CAsT'19, user utterances may only refer to the information mentioned in previous user utterances.  Since 2020~\citep{Dalton:2020:TREC}, utterances may refer to previous responses given by the system as well, which significantly extends the scope of contextual information that the system needs to use to understand a request. TREC CAsT'21~\citep{Dalton:2021:TREC} is characterized by the increased dependence on previous system responses, as well as simple forms of user revealment, reformulation, and explicit feedback introduced in users' utterances. 

By TREC CAsT'21, a two-step passage ranking architecture has emerged.  A first-pass passage retrieval is usually performed using an unsupervised sparse model (e.g., BM25), which is followed by re-ranking using a neural model trained for passage retrieval (e.g., T5 trained on MS MARCO~\citep{Craswell:2020:TREC}).  Additionally, most systems employ a query rewriting step, where the original query is de-contextuali\-zed to be independent of the previous turns. 

\subsection{Query Rewriting}
\label{sec:related_works:query_rewriting}

The goal of query rewriting is to handle common conversational phenomena such as omission, coreference~\citep{Dalton:2019:TREC}, zero anaphora, topic change, and topic return~\citep{Voskarides:2020:SIGIR}.
Approaches can be broadly categorized into unsupervised, supervised feature-based, and (weakly-)supervised neural methods.
Unsupervised query rewriting methods expand the original query with terms from the conversation history, for example, from previous utterances based on BM25 score~\citep{Yilmaz:2019:TREC}, cosine similarity~\citep{Voskarides:2019:TREC}, or other frequency-based signals~\citep{Lin:2021:TOIS}.
Supervised feature-based methods use linguistic features based on dependency parsing, coreference resolution, named entity resolution, or part-of-speech tagging~\citep{Mele:2020:SIGIR}.
Supervised neural query rewriting approaches utilize large pre-trained language models, and in particular generative models such as GPT-2~\citep{Vakulenko:2021:ECIR} or T5 model~\citep{Yan:2021:TREC,Ju:2021:TREC,Chang:2020:TREC}. These models are fine-tuned on a conversational dataset, such as CANARD~\citep{Vakulenko:2021:ECIR, Lin:2021:TOIS, Ju:2021:TREC, Chang:2020:TREC, Vakulenko:2021:WSDM} or QReCC~\citep{Yan:2021:TREC}.  The generated query reformulations may further be expanded with terms from conversation history~\citep{Vakulenko:2021:ECIR}, with paraphrases~\citep{Ju:2021:TREC}, or related sentences from semantically related documents~\citep{Chang:2020:TREC}. 
Weakly supervised neural query rewriting methods aim to fine-tune large pre-trained language models~\citep{Yu:2020:SIGIR} or term selection classifiers~\citep{Kumar:2020:EMNLP} with weak supervision data that is created using rule-based or self-supervised approaches.
The best results are reported using a combination of term-based query expansion with generative models for query reformulation~\citep{Kumar:2020:EMNLP,Lin:2021:TOIS,Vakulenko:2021:ECIR}.

\begin{figure}[t]
\begin{tabular}{p{4.1cm}p{3.5cm}p{3.6cm}}
    \hspace*{-1mm}\includegraphics[width=42mm]{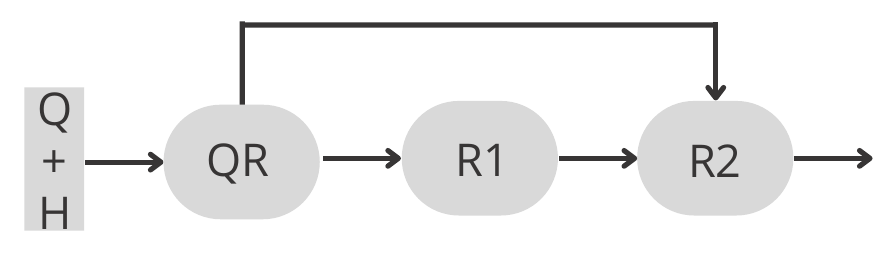} &  
    \includegraphics[width=35mm]{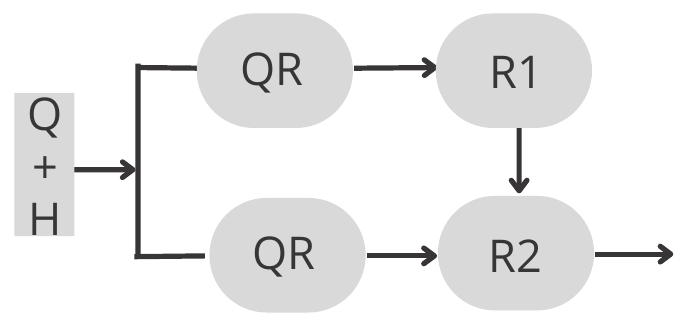} &
    \hspace*{-1mm}\includegraphics[width=38mm]{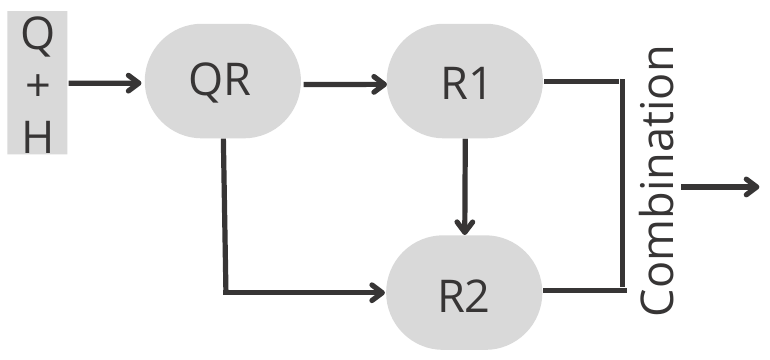}\\
    \scriptsize{(a\speciallabel{a}{same_qr}) Basic two-stage retrieval pipeline using a single query rewriter.} & 
    \scriptsize{(b\speciallabel{b}{different_qr}) Different query rewriter for first-pass retrieval and re-ranking.} & 
    \scriptsize{(c\speciallabel{c}{combination}) Combination of first-pass retrieval and re-ranking using the same query rewriting.}
     \\[6pt]
    \hspace*{-2mm}\includegraphics[width=44mm]{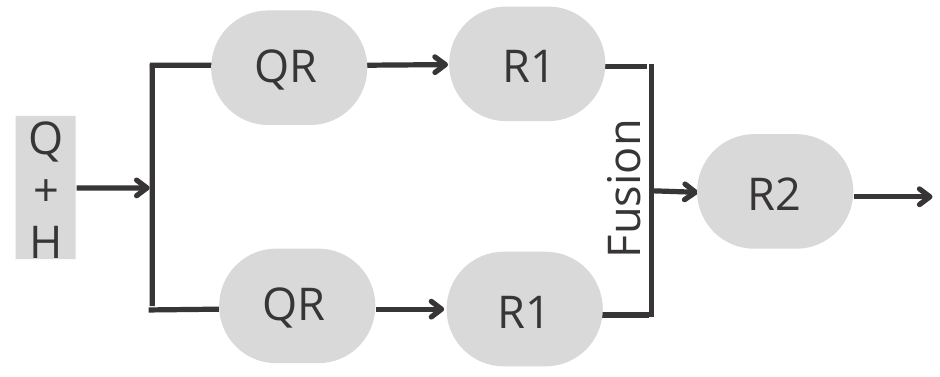} &
    \hspace*{-1mm}\includegraphics[width=37mm]{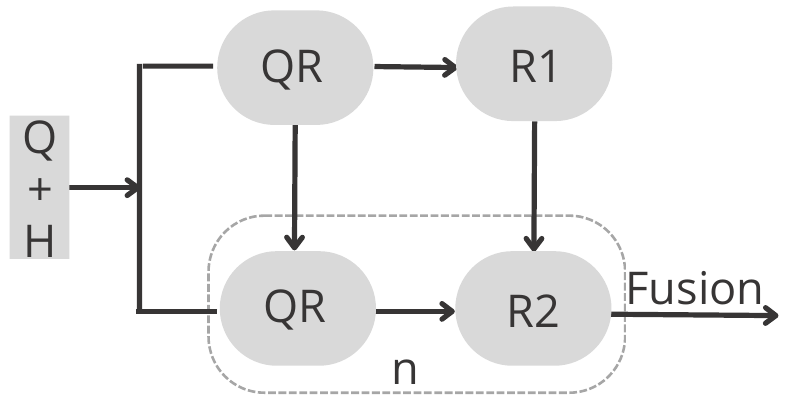} &   \hspace*{-1mm}\includegraphics[width=37mm]{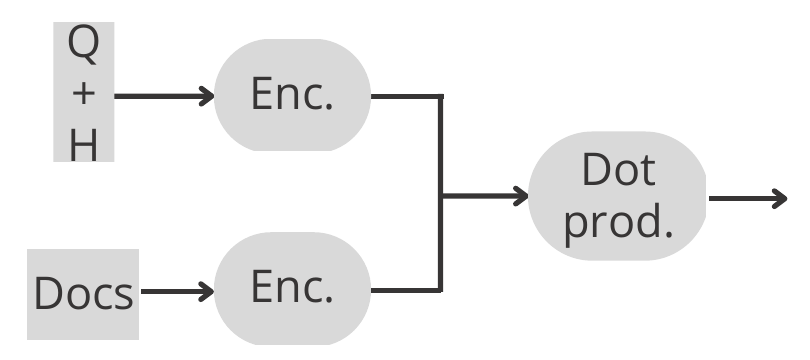} \\
    \scriptsize{(d\speciallabel{d}{fusion})  Re-ranking of fused first-pass results that use different query rewriters.} & 
    \scriptsize{(e\speciallabel{e}{n_fusion}) Fusion of multiple passage re-rankings using different rewrites.} & 
    \scriptsize{(f\speciallabel{f}{few-shot})  Few-shot conversational dense retrieval.} \\[6pt]
\end{tabular}
\caption{Pipeline architectures for conversational search (\textbf{Q+H}: raw query and conversational history; \textbf{QR}: query rewriter; \textbf{R1}: first-pass retriever; \textbf{R2}: re-ranker; \textbf{Enc.}: encoder; \textbf{Docs}: document collection; \textbf{Dot prod.}: dot product).}
\label{fig:architectures}
\end{figure}

\subsection{Pipeline Architectures}

Systems participating in TREC CAsT exhibit a wide variety of approaches, not only in terms of component-level choices but also in terms of the overall architectures of their ranking pipelines.  The most common choice is a two-stage retrieval pipeline with a query rewriting module. Different variants of this cascade architecture include systems with the same rewriting method used for both first-pass retrieval and re-ranking~\citep{Chang:2020:TREC, Vakulenko:2020:TREC, Yan:2021:TREC, Ju:2021:TREC, Vakulenko:2021:ECIR, Yu:2020:SIGIR, Vakulenko:2021:WSDM, Mele:2020:SIGIR} (Fig.~\ref{fig:architectures}\ref{same_qr}), different query rewriting modules for both stages~\citep{Yang:2019:TREC} (Fig.~\ref{fig:architectures}\ref{different_qr}), or using rewriting only for first-pass retrieval~\citep{Gemmell:2020:TREC, Yang:2019:TREC}. 

More advanced architectures may use a two-stage retrieval pipeline with the same query rewriter for each stage, but combine the scores obtained from retrieval and re-ranking to produce a final ranking~\citep{Voskarides:2019:TREC} (Fig.~\ref{fig:architectures}\ref{combination}) or use two different versions of the rewritten query for first-pass retrieval and a fusion of the ranked lists for the re-ranking stage~\citep{Lin:2021:TOIS} (Fig.~\ref{fig:architectures}\ref{fusion}). Another architecture variant consists of first-pass retrieval using the rewritten query, followed by a fusion of multiple contextualized passage re-ranking of several different rewrites~\citep{Kumar:2020:EMNLP} (Fig.~\ref{fig:architectures}\ref{n_fusion}).
An alternative to the retrieve-then-rerank approach is a few-shot conversational dense retrieval system that learns contextualized embeddings of utterances and documents in the collection, and scores documents solely using the dot product of the embeddings~\cite{Yu:2021:SIGIR} (Figure~\ref{fig:architectures}\ref{few-shot}).

\section{Selected Approaches}
\label{sec:reproduced_approaches}

We present the two approaches from TREC CAsT'21 that we aim to reproduce in this paper: (1) the best performing official baseline provided by the track organizers' and (2) the top performing system submitted by participants.\footnote{More specifically, this is the best performing system that is accompanied by a system description and can thus be (attempted to be) reproduced.}
These approaches may be regarded as representatives of a strong baseline and of the state of the art, respectively.
Both may be seen as instantiations of the basic two-stage retrieval pipeline approach (cf. Fig.~\ref{fig:architectures}\ref{same_qr}), with query rewriting, first-pass retrieval, and re-ranking components, as shown in Table~\ref{tab:overview_reproduced_approaches}.
In this section, we focus on a high-level description of these approaches, based on the corresponding TREC papers; specific implementation details are discussed in Section~\ref{sec:reproducibility_experiments}.

\begin{table*}[t]
    \scriptsize
    \caption{Overview of approaches reproduced in this paper.}
    \label{tab:overview_reproduced_approaches}
    \begin{tabular}{p{2.3cm}|p{3.4cm}|p{3.2cm}|p{1.6cm}}
        \hline
        & \textbf{Query rewriting} & \textbf{First-pass retrieval} & \textbf{Re-ranking} \\
        \hline
        BaselineOrganizers & T5 fine-tuned on CANARD & BM25 & monoT5 \\ 
        \hline
        WaterlooClarke  & T5 fine-tuned on QReCC & BM25 with PRF + ANCE & mono/duoT5 \\ 
        \hline
    \end{tabular}
\end{table*}

\subsection{Organizers' Baseline}
\label{sec:reproduced_approaches:baseline}

Of the several baselines provided by the track organizers, \texttt{org\_auto\_bm25\_t5} was the best performing run~\citep{Dalton:2021:TREC}; this will be referred to as the \textbf{BaselineOrganizers} approach henceforth. The query rewriting component is using T5 fine-tuned on CANARD for generative query rewriting. The rewriter uses all previous queries and the three previous canonical responses as context. For first-pass retrieval, BM25 is used to collect the top 1000 documents from the collection. The documents are re-ranked with a pointwise (mono) T5 model trained on MS MARCO.

\subsection{Top Performer: WaterlooClarke}
\label{sec:reproduced_approaches:waterlooclarke}

The top-performing documented system was the \texttt{clarke-cc} run by~\citet{Yan:2021:TREC}; this will be referred to as the \textbf{WaterlooClarke} approach henceforth.
The query rewriting component is based on a T5 model that is fine-tuned on the QReCC dataset~\citep{Anantha:2021:NAACL}. For context, the rewriter uses previously rewritten utterances and the last canonical result.
First-pass retrieval comprises two sub-components: a sparse and a dense retriever.  The sparse retriever utilizes a BM25 with pseudo-relevance feedback (PRF), with the parameters tuned to maximize recall. PRF is run over both the target corpus and the C4 corpus.\footnote{ \url{https://huggingface.co/datasets/allenai/c4}}  The dense retriever is based on the ANCE approach~\citep{Xiong:2020:ICLR}. Both retrieval systems return the top 1000 documents that are merged into one final ranking.
Re-ranking is performed using a pointwise T5 re-ranker, followed by another re-ranking of the top 50 documents, using pairwise duoT5~\citep{Pradeep:2021:arXiv}. 

\section{Reproducibility Experiments}
\label{sec:reproducibility_experiments}

In this section, we answer our first research question: Can the organizers' baseline and the best performing system at the TREC CAsT'21 be reproduced? 
We describe the implementation details of the two systems and discuss their end-to-end performance with respect to the results reported in the track overview~\citep{Dalton:2021:TREC}.

\subsection{Baseline Implementation}
\label{sec:reproducibility:baseline}

We base the implementation solely on the description of the track organizers' \texttt{org\_auto\_bm25\_t5} baseline in the overview paper~\citep{Dalton:2021:TREC}, without resorting to additional communication with the authors.

The passage collection is indexed using Elasticsearch, using the built-in analyzer for tokenization, stopwords removal, and KStem stemming. 
For query rewriting, we use a pre-existing T5 model that has been fine-tuned on the CANARD dataset (\texttt{castorini/t5-base-canard}).\footnote{\url{https://huggingface.co/castorini/t5-base-canard}}  Our implementation is based on the Hugging Face transformers library.\footnote{\url{https://github.com/huggingface/transformers}} 
According to~\citep{Dalton:2021:TREC}, the context for the query rewriter is of the form: $q_1, q_2, \dots, q_{i-3}, r_{i-3}, q_{i-2}, r_{i-2}, q_{i-1}, r_{i-1}, q_i$, where $q_i$ and $r_i$ are the \emph{i}th raw query and canonical response, respectively. Contexts exceeding the allowed model input length are not handled. This, however, can result in trimming the input in a way that the raw query that is to be rewritten is removed. To increase the quality of the rewriting by ensuring the correct form of the input and benefiting from previous rewrites, we alternatively use: $\hat{q}_1, \hat{q}_2, \dots , \hat{q}_{i-1}, trim(r_{i-1}), q_i$, where $\hat{q}_{i}$ is the \emph{i}th rewritten query and \emph{trim} is a function that cuts the canonical response if the length of the input is longer than the capacity of the model. 
For first-pass retrieval, the passages are ranked using BM25 on a \texttt{catch\_all} field (concatenating the \texttt{title} and \texttt{body} fields) in the 2021 index and on the \texttt{body} field for the 2020 index.  We initially used the parameters reported by the organizers (k1=4.46, b=0.82), but then achieved better results with the default parameters (k1=1.2, b=0.75).
The top 1000 candidates for each turn are re-ranked using the T5 model introduced by~\citet{Nogueira:2020:EMNLP}, which has been published on Hugging Face (\texttt{castorini/monot5-base-msmarco}).\footnote{\url{https://huggingface.co/castorini/monot5-base-msmarco}}

\subsection{WaterlooClarke Implementation}

We base our implementation on the WaterlooClarke group's TREC paper~\citep{Yan:2021:TREC}.  Additional information on specific details was obtained from the authors via email communication and inferred from the implementation made available.\footnote{\url{https://github.com/claclark/Cottontail/blob/main/apps/treccast21.cc}}

The approach requires two indices: an approximate nearest neighbor (ANN) index for ANCE dense retrieval and an inverted index for BM25. The authors use ANCE's own implementation\footnote{\url{https://github.com/microsoft/ANCE}} and a publicly released model checkpoint (passage ANCE(FirstP)) for the ANN index.\footnote[$\dagger$]{\label{personal_communication}Missing information provided by the authors in personal communication.} We use Pyterrier's plugin\footnote{\label{pyterrier_ance}\url{https://github.com/terrierteam/pyterrier_ance}} for creating the ANN index, which is based on the original paper, and allows for easier integration with other modules in our pipeline.
For building the ANN index we use MS MARCO Passage and TREC CAR collections provided by the ir\_datasets package,\footnote{\url{https://github.com/allenai/ir_datasets}} and implement our own generator for the WaPo 2020, MS MARCO Documents, and KILT collections. No additional preprocessing is performed when building the dense retrieval index. The inverted index used by BM25 is the same as in Section~\ref{sec:reproducibility:baseline}.

The query reformulation step in WaterlooClarke is based on a T5 model trained on the QReCC dataset~\citep{Anantha:2021:NAACL}. All the previous rewritten utterances and the canonical response for the last utterance are used as context to reformulate the current question (i.e., the input is given as: $\hat{q}_1, \hat{q}_2, \dots, \hat{q}_{i-1}, trim(r_{i-1}), q_i$). If the length of the input sentence exceeds 512, the answer passage is cut off.\textsuperscript{\ref{personal_communication}} The authors fine-tune a pretrained \texttt{t5-base} model\footnote{\url{https://huggingface.co/t5-base}} with the training partition of the QReCC dataset for 3 epochs, using the original test partition as a validation set.\textsuperscript{\ref{personal_communication}} The train batch size is equal to 2 and the learning rate is $5 \times 10^{-5}$.\textsuperscript{\ref{personal_communication}} 
We use the Simple Transformers library\footnote{\url{https://simpletransformers.ai/}} for the fine-tuning procedure (as opposed to PyTorch Lightning\footnote{\url{https://www.pytorchlightning.ai/}} and Hugging Face transformers used by the authors\textsuperscript{\ref{personal_communication}}). 

There are two first-pass rankers involved: (1) sparse retrieval using BM25 with pseudo relevance feedback (PRF) and (2) dense retrieval using ANCE~\citep{Xiong:2020:ICLR}. 
The final sparse retrieval ranking is a fusion of two rankings.\textsuperscript{\ref{personal_communication}} PRF is applied on the top 17 documents to expand the query with the top 26 terms; the expanded query is then scored using BM25 to generate the first sparse ranking. Additionally, the authors use the top 16 weighted answer candidates generated by a statistical question-answering method ran against the C4 corpus to create the second ranking (answer candidates are used by BM25).\textsuperscript{\ref{personal_communication}} The first and the second ranking produced by the sparse retrieval are fused with Reciprocal Rank Fusion (RRF)~\citep{Cormack:2009:SIGIR}.\textsuperscript{\ref{personal_communication}} 
There is no further information disclosed about the question-answering system used (neither in the paper nor in the GitHub repository). Therefore, we skip the second ranking in reproducibility and focus on standard BM25 with PRF.  
The BM25 parameters are tuned to maximize recall over manually rewritten questions from previous years.  The exact details of this remain unclear. 
We tune BM25 parameters on our 2020 and 2021 indices and take the average of the best parameters found for each year (b=0.45, k1=0.95), since the parameters used in their code (b=0.45, k1=1.18) gave worse results on our indices. For query expansion, since the choice of PRF algorithm could not be resolved, we opted for RM3~\citep{Lavrenko:2001:SIGIR}, which we implemented from scratch.

The results of sparse and dense retrieval are fused to generate the final set of 1000 candidate passages for re-ranking. Since the fusion method is not stated in the paper, we assume that this step also employs RRF; we utilize the TrecTools library,\footnote{\url{https://github.com/joaopalotti/trectools}} which implements a RRF as defined in~\citep{Cormack:2009:SIGIR}.

The re-ranking stage in this approach is based on a pointwise monoT5 re-ranker (on all candidate passages), followed by a pairwise duoT5 re-ranker (on the top 50 passages re-ranked by monoT5). The original re-ranking implementation is based on the Pyaggle library\footnote{\url{https://github.com/castorini/pygaggle}} with the default model checkpoints.  Our implementation of duoT5 is based on the Hugging Face transformers library and the  \texttt{castorini/duot5-base-msmarco} model published on Hugging Face.\footnote{\url{https://huggingface.co/castorini/duo5-base-msmarco}}

\subsection{Results}
\label{sec:reproducibility_experiments:results}
  
Table~\ref{tab:reproducibility:performance} reports our results on the CAsT'21 collection.
Following the official setup, we consider measures with both binary and graded relevance. The main measure is NDCG@3; other measures are computed with a rank cutoff of 500. For binary measures, we apply a relevance threshold of 2.

\begin{table*}[t]
    \scriptsize
	\caption{Reproducibility experiments on the TREC CAsT'21 dataset.} \label{tab:reproducibility:performance}
	\begin{tabular}{p{5cm}|p{0.9cm}|p{0.9cm}|p{0.9cm}|p{0.9cm}|p{1.3cm}}
		\hline
		\textbf{Approach} & \textbf{R@500} & \textbf{MAP} &
        \textbf{MRR} & \textbf{NDCG} & \textbf{NDCG@3} \\
		\hline
		BaselineOrganizers@TREC'21 (in~\citep{Dalton:2021:TREC}) & 0.636 & 0.291 & 0.607 & 0.504 & 0.436 \\ 
		BaselineOrganizers-QR-BM25 & 0.5632 & 0.2268 & 0.4947 & 0.4317 & 0.3457 \\ 
		BaselineOrganizers-BM25 & 0.5894 & 0.2546 & 0.5405 & 0.4672 & 0.3966 \\ 
		BaselineOrganizers & 0.6472 & 0.2628 & 0.5354 & 0.4885 & 0.3968 \\ 
		\hline
		WaterlooClarke@TREC'21 (in~\citep{Dalton:2021:TREC}) & 0.869 & 0.362 & 0.684 & 0.640 & 0.514 \\ 
		WaterlooClarke@TREC'21 (runfile) & 0.8534 & 0.3494 & 0.6626 & 0.6240 & 0.4950 \\ 
		WaterlooClarke reproduced by us & 0.6915 & 0.2864 & 0.5712 & 0.5176 & 0.4151 \\
		\hline
	\end{tabular}
\end{table*}

For the baseline, the results reported in the overview paper~\citep{Dalton:2021:TREC} are included verbatim and regarded as the reference, since the raw runfile (\texttt{org\_auto\_bm25\_t5}) is not available in the TREC archive.
We include results using the original query rewriting method and reported BM25 parameters (BaselineOrganizers-QR-BM25), using the improved query rewriter while keeping the reported BM25 parameters (BaselineOrganizers-BM25), and finally using the improved query rewriter with default BM25 parameters (BaselineOrganizers).  We find that the latest variant performs best; it is still 9\% below the reference result in terms of NDCG@3, but 2\% better in terms of Recall@500.

Regarding WaterlooClarke, the performance of our reproduced system is 19\% lower in terms of NDCG@3 and 20\% lower in terms of Recall@500 than the official results reported for this approach.  
The discrepancy in the results is most likely caused by the lack of the C4-based question-answering step performed in first-pass retrieval. This element of the system is not sufficiently described in the paper nor has been resolved via personal email communication.  
Surprisingly, we observe discrepancies between the official results reported in the overview paper and a direct evaluation of the \texttt{clarke-cc} runfile taken from the TREC archive (cf. rows 5 vs. 6 in Table~\ref{tab:reproducibility:performance}).
The latter results are lower, with a relative drop of almost 4\% in NDCG@3, which is a non-negligible difference.  We cannot explain this discrepancy; however, it also puts into question the results reported in the track overview.
When comparing our reproduced results against their runfile, the relative differences are under 16\% and 19\% in terms of NDCG@3 and Recall@500, respectively.

Overall, according to the track overview paper, the relative differences between BaselineOrganizers and WaterlooClarke are 18\% and 37\% in terms of NDCG@3 and Recall@500, respectively (cf. rows 1 vs. 5 in Table~\ref{tab:reproducibility:performance}).  However, the respective differences in our reproduced approaches are 5\% and 7\% (cf. rows 4 vs. 7 in Table~\ref{tab:reproducibility:performance}).  Moreover, these differences are no longer statistically significant, based on a paired t-test with $p<0.05$. The same test does indicate significant differences when performed against the WaterlooClarke runfile.

\subsection{Summary}

In summary, neither approach could be fully reproduced due to key information missing.
In the case of BaselineOrganizers, the specifics of the models used for query rewriting and re-ranking were lacking, and the formulation of input sequences for query rewriting was underspecified (esp. with regards to exceeding the length limits of the model).
As for WaterlooClarke, the complexity of the system and shortages in technical details made it impossible to fully implement the system.  Most notably, the involvement of a question-answering system for sparse retrieval is not even mentioned in the paper.  We do want to acknowledge the kind, helpful, and open communication by the authors via email, which allowed us to resolve questions around the query rewriting model and its parameters, the BM25 and PRF parameters used, and the rank fusion method employed.  Nevertheless, after several rounds of email exchanges, we are still missing details about the PRF algorithm, the question-answering system employed, the exact approach used for tuning the BM25 parameters, the preprocessing employed for the inverted index, and the method used for combining sparse and dense rankings.
It is also worth noting that while BM25 parameters were shared for both approaches, those parameters were not the optimal ones for us, which is likely due to differences in document preprocessing. It, however, means that BM25 parameters alone, without further details on preprocessing or collection statistics, are only moderately useful.
We shall reflect more generally on some of these challenges and possible remedies in Section~\ref{sec:concl}.

\section{Additional Experiments}
\label{sec:additional}

We have reproduced two approaches, BaselineOrganizers and WaterlooClarke, which follow the same basic two-stage retrieval pipeline (cf. Fig.~\ref{fig:architectures}\ref{same_qr}), but differ in each of the query rewriting, first-pass retrieval, and re-ranking components.  
We experiment with different configurations of this basic pipeline to understand which changes contribute most to overall performance (Section~\ref{sec:additional:qr}).
Additionally, we consider a different pipeline architecture (Section~\ref{sec:additional:architecture}).
In both sets of experiments, we are interested in the generalizability of findings, therefore we also report results on the TREC CAsT'20 dataset. (Note that the rank cut-off for 2020 collection is 1000, while for 2021 it is 500.)

\begin{table*}[t]
    \tiny
	\caption{Variants of a two-stage retrieval pipeline on TREC CAsT'20 and '21.} \label{tab:query_rewriting:performance}
	\begin{tabular}{p{5.4cm}|p{0.9cm}|p{0.7cm}|p{0.7cm}|p{0.7cm}|p{1.1cm}}
	    \hline
		\textbf{Approach} & \textbf{R@1000} & \textbf{MAP} &
        \textbf{MRR} & \textbf{NDCG} & \textbf{NDCG@3} \\
		\hline
		\multicolumn{6}{c}{\textbf{TREC CAsT 2020}} \\
		\hline
		T5\_CANARD + BM25 + monoT5 & 0.5276 & 0.2191 & 0.5457 & 0.4353 & 0.3789 \\
		T5\_QReCC + BM25 + monoT5 & 0.5100 & 0.2056 & 0.5106 & 0.4065 & 0.3618 \\
        T5\_CANARD + ANCE/BM25 + mono/duoT5 & 0.6781 & 0.2540 & 0.5512 & 0.5027 & 0.4052 \\
		T5\_QReCC + ANCE/BM25 + mono/duoT5 & 0.6449 & 0.2443 & 0.5357 & 0.4804 & 0.4061 \\
		T5\_CANARD + ANCE/BM25/PRF + mono/duoT5 & \textbf{0.6878} & \textbf{0.2555} & \textbf{0.5541} & \textbf{0.5063} & \textbf{0.4086} \\
		T5\_QReCC + ANCE/BM25/PRF + mono/duoT5 & 0.6608 & 0.2451 & 0.5355 & 0.4840 & 0.4052 \\ 
		\hline
		\textbf{Approach} & \textbf{R@500} & \textbf{MAP} &
        \textbf{MRR} & \textbf{NDCG} & \textbf{NDCG@3} \\
		\hline
		\multicolumn{6}{c}{\textbf{TREC CAsT 2021}} \\
		\hline
		T5\_CANARD + BM25 + monoT5 & 0.6472 & 0.2628 & 0.5354 & 0.4885 & 0.3968 \\
		T5\_QReCC + BM25 + monoT5 & 0.6018 & 0.2530 & 0.5369 & 0.4670 & 0.3933 \\
		T5\_CANARD + ANCE/BM25 + mono/duoT5 & 0.7259 & 0.2886 & 0.5575 & 0.5316 & 0.4068 \\
		T5\_QReCC + ANCE/BM25 + mono/duoT5 & 0.6799 & 0.2843 & 0.5702 & 0.5135 & \textbf{0.4159} \\
		T5\_CANARD + ANCE/BM25/PRF + mono/duoT5 & \textbf{0.7306} & \textbf{0.2915} & 0.5573 & \textbf{0.5330} & 0.4061 \\
		T5\_QReCC + ANCE/BM25/PRF + mono/duoT5 & 0.6915 & 0.2864 & \textbf{0.5712} & 0.5176 & 0.4151 \\
		\hline
	\end{tabular}
\end{table*}

\subsection{Variants of a Two-Stage Retrieval Pipeline}
\label{sec:additional:qr}

In this experiment, we gradually switch out the components of a baseline system (BaselineOrganizers) with components of a state-of-the-art system (WaterlooClarke).
The results are presented in Table~\ref{tab:query_rewriting:performance}; the first and last rows within each block correspond to BaselineOrganizers and WaterlooClarke, respectively.
Our observations are as follows.
First, when changing the dataset used for training the T5-based query rewriter from CANARD to QReCC (rows 1 vs. 2, 3 vs. 4, and 5 vs. 6 in Table~\ref{tab:query_rewriting:performance}) we observe a noticable drop (3\%--7\%) in terms of recall, with smaller differences in NDCG@3 (below 2\%, with one exception).
Second, using more advanced retrieval methods (ANCE/BM25 instead of BM25 for first-pass ranking and mono/duoT5 instead of monoT5 for re-ranking; rows 1 vs. 3 and 2 vs. 4 in Table~\ref{tab:query_rewriting:performance}) does yield consistent improvements across metrics and datasets: +12\%--29\% in recall and +3\%--12\% in NDCG@3.
Finally, using pseudo relevance feedback for first-pass retrieval (rows 3 vs. 5 and 4 vs. 6 in Table~\ref{tab:query_rewriting:performance}) results in small but consistent improvements in terms of recall (1\%--2\%) with negligible differences in NDCG@3 ($<$1\%).
It should be noted that none of the above differences are statistically significant, thereby the results are merely indicative.
However, in terms of overall trends, our results are in line with the tendencies reported by~\citet{Yan:2021:TREC}. Namely, that adding PRF and combining sparse and dense retrieval methods for first-pass retrieval improves performance.

\begin{table*}[t]
    \scriptsize
	\caption{Performance of query rewriting approaches with different variants of the two-stage pipeline on the TREC CAsT'20 and '21 datasets. Highest scores for each year are in boldface.} \label{tab:architecture:performance}
	\centering
	\begin{tabular}{ |c|c|c|c|c| } 
    \cline{2-5}
    \multicolumn{1}{c|}{} & Recall & NDCG@3 & Recall & NDCG@3 \\ 
    \hline
    \diagbox{R1}{R2} & \multicolumn{2}{c|}{T5\_CANARD} & \multicolumn{2}{c|}{T5\_QReCC} \\
    \hline
    \multirow{2}{*}{T5\_CANARD} & 2020: \textbf{0.6878}  & 2020: \textbf{0.4086} & 2020: \textbf{0.6878} & 2020: 0.3923 \\ 
    & 2021: \textbf{0.7306} & 2021: 0.4061 & 2021: 0.7267 & 2021: 0.4166 \\ 
    \hline
    \multirow{2}{*}{T5\_QReCC} & 2020: 0.6608 & 2020: \textbf{0.4086} & 2020: 0.6608 & 2020: 0.4052 \\ 
    & 2021: 0.6879 & 2021: \textbf{0.4176} & 2021: 0.6915 & 2021: 0.4151 \\ 
    \hline
    \end{tabular}
\end{table*}

\subsection{Using a Different Pipeline Architecture}
\label{sec:additional:architecture}

It is clear that query rewriting has a direct impact on both ranking steps: first-pass retrieval (R1) and re-ranking (R2).  Still, it remains to be seen whether the two stages are impacted the same way.
The basic two-stage retrieval pipeline (cf. Fig.~\ref{fig:architectures}\ref{same_qr}) uses the same query rewriter for both ranking stages and therefore cannot be used to answer this question.  We thus switch to a different pipeline architecture---one that uses a different query rewriter component for R1 and R2, but is identical to the basic pipeline in the ranking components (cf. Fig.~\ref{fig:architectures}\ref{different_qr}).

Table~\ref{tab:architecture:performance} presents the results for the possible four-way combinations of query rewriters, T5\_CANARD and T5\_QReCC, and ranking stages, R1 and R2.  The ranking components follow the WaterlooClarke approach (i.e., using T5\_QReCC for both R1 and R2 corresponds to the last row in Table~\ref{tab:query_rewriting:performance}).
The results reveal some interesting tendencies that generalize across both datasets (even though the differences are not statistically significant). Using T5\_CANARD for first-pass retrieval results in the highest recall. However, the overall best combination in terms of final ranking (NDCG@3) is when T5\_QReCC is employed in first-pass retrieval and T5\_CANARD is used in re-ranking. 
Overall, we observe meaningful relative improvements for recall (up to 6\%) and negligible improvements for NDCG@3 ($\leq$1\%) on both datasets over the WaterlooClarke approach.

\section{Reflections and Conclusions}
\label{sec:concl}

In this work, we have attempted to reproduce approaches for the task of conversational passage retrieval, in the context of TREC CAsT.
TREC papers can range anywhere between vague system descriptions to full-fledged research papers, which can make reproducibility a real challenge; this has certainly been the case for this study.
We acknowledge that reproducibility is not a requirement for TREC submissions.  Still, since they are often used for reference comparison in terms of absolute system performance on a given test collection (cf.~\citep{Armstrong:2009:CIKM}), it is worth considering how easy or difficult it is to reproduce them.  Specifically, we have selected two approaches for our study: the best performing baseline by the track organizers and the best performing participant submission (that was accompanied by a paper) from the 2021 edition of TREC CAsT.
We have decided against personal communication with the track organizers (thus implicitly subjecting them to a higher virtual bar-of-standard) while making a best effort to resolve any missing details with the participant team over email.

Overall, our reproducibility efforts have met with moderate success.  Surprisingly, we have managed to come closer to reproducing the participant's submission (WaterlooClarke) than the organizers' baseline.  In the case of the former, there is a missing sparse retrieval component that can well explain the difference. As for the organizers' results, the discrepancies between the reported results in the track overview paper and the actual runfiles found in the TREC archive would be worth a follow-up investigation.
Generally, key missing information includes the names of specific algorithms and models used, and detailed-enough descriptions of procedures of constructing inputs to neural models and ways of obtaining models' parameters.  We wish to note that sharing model parameters in some cases is not enough; consider, e.g., the simple case of BM25, where the length normalization parameter alone is not meaningful if collection statistics markedly differ due to how the collection is preprocessed.
Given that multi-stage ranking architectures are common at TREC CAsT, but also beyond that, sharing intermediate results from the different components would be immensely valuable. These could include the rewritten or expanded queries, set of candidate document IDs, and intermediate document rankings.  Sharing them would not only support reproducibility but also facilitate component-level evaluation.

Since the two reproduced systems follow the same basic two-stage retrieval pipeline, we have also performed additional experiments to study different configurations of this pipeline and have made some observations regarding the contributions made by the various components.  Moreover, we have reported on experiments with different combinations of query rewriting methods using a different retrieval pipeline, which have yielded some novel findings.  Further comparisons of different pipeline architectures would be an especially interesting direction for future work.

\paragraph{Post-acceptance communication with TREC CAsT organizers.}
Upon acceptance of this paper, we attempted to clarify the discrepancies between the results in this paper and those reported in the track overview via email communication with the track organizers. 
There is a difference in tooling: they used Pyserini\footnote{\url{https://github.com/castorini/pyserini}} for building the index, while we used Elasticsearch.  Differences in collection preprocessing (tokenization, stemming, stopword removal, etc.) may contribute to the gap in the results. 
Regarding the runfile, we were pointed to the track's GitHub repository\footnote{\url{https://github.com/daltonj/treccastweb/tree/master/2021/baselines}} containing the raw runfile (\texttt{org\_automatic\_results\_1000.v1.0.run}). 
However, evaluating this runfile against the official qrels still yields results different from those reported in the track overview paper (in parentheses): Recall@500 is 0.623 (vs. 0.636), MAP is 0.282 (vs. 0.291), and NDCG@3 is 0.424 (vs. 0.436).
This is ``in alignment'' with the case of the WaterlooClarke (\texttt{clarke-cc}) runfile, in the sense that there is a mismatch between the numbers reported in the track overview paper and the evaluation of the actual runfiles (with the latter being lower).
At the time of writing, this issue has not been resolved.  We plan to update our online repository if new findings become available.

\subsubsection{Acknowledgments}

This research was supported by the Norwegian Research Center for AI Innovation, NorwAI (Research Council of Norway, project number 309834).  We thank the members of the WaterlooClarke group (School of Computer Science, University of Waterloo, Canada), Xinyi Yan and Charlie Clarke for supporting our efforts to reproduce their TREC CAsT'21 submission. We also thank the TREC CAsT organizers for their efforts in coordinating the track and for providing us with additional technical details regarding their baseline.

\renewcommand*{\bibfont}{\scriptsize}
\bibliographystyle{abbrvnat}
\bibliography{ecir2023-treccast.bib}

\end{document}